\documentstyle[eqsecnum,aps]{revtex}
\begin{document}

\date{}
\title{A simple formula for ground state energy of a two-electron atom}
\author{M. Auzinsh, R. Damburg \\
Dept. of Physics, University of Latvia\\
19. Rainis blvd., Riga LV-1586, Latvia}
\maketitle

\begin{abstract}
A simple expression for a ground state energy for a two-electron atom is
derived. For this assumption based upon the Niels Bohr ''old'' quantum
mechanics idea about electron correlation in a two-electron atom is
exploited. Results are compared with experimental data and theoretical
results based on a variation approach.
\end{abstract}

\section{Introduction}

The helium problem played an important role in the early history of quantum
mechanics \cite{hy3}. Already at the very beginning of ''old'' quantum
theory Niels Bohr made his first attempt to calculate energy levels of the
He atom. He discussed a model when both electrons of a two-electron atom
move along the same circular orbit and are located at the opposite ends of a
diameter \cite{bo1}. Even some time before Bohr's attempt Hantaro Nagaoka in
Japan demonstrated from the view point of classical mechanics that such
motion possesses the lowest possible energy and -- what is very important --
it is a mechanically stable arrangement \cite{na1}. It means that weak
disturbances cannot destroy this motion.

Afterwards Bohr's approach to the problem was considered to be unsuccessful
mainly due to the reason that this model fails to the explain diamagnetism
of a two-electron atom in its lowest energy state and also the value of the
obtained energy was in a rather poor agreement with experiment. Later it was
assumed that only quantum mechanics can satisfactorily explain such a
two-electron atom.

In this paper we will try to demonstrate, that if we combine quantum
mechanical approach to the two-electron atom with ideas that were at the
heart of Bohr's approach, we can obtain a very simple expression for the
lowest energy level of a two-electron atom that can nevertheless describe
these energies with the same accuracy as do more complicated approaches
based on the quantum variation method combined with a quantum perturbation
theory.

\section{Schroedinger equation for a two-electron atom with strong electron
correlation}

Let us consider a two-electron atom with a nuclear charge $Z.$ In atomic
units the stationary Schroedinger equation for this system is 
\begin{equation}
\left( -\frac 12\Delta _1-\frac 12\Delta _2-\frac Z{r_1}-\frac Z{r_2}+\frac
1{r_{12}}\right) \psi =E\psi .  \label{eq1}
\end{equation}
It is well known that to solve this equation, it is important to account
correctly for electron correlation. Now let us assume, as it was done by
Niels Bohr, that electrons are permanently located exactly on the opposite
sides of nucleus, so that \textbf{\ }$\mathbf{r}_1=-a\mathbf{r}_2$ ($a$
being positive). In opposition to the general case of the Schroedinger
equation for a two-electron atom, with this last assumption, equation $%
\left( \ref{eq1}\right) $ can be solved precisely analytically. For
eigenvalues the solution is 
\begin{equation}
E_n\left( a,Z\right) =-\frac{\left( 1+a\right) ^2}{2n^2(1+a^2)}\left(
Z-\frac a{(1+a)^2}\right) ^2,  \label{eq2}
\end{equation}
where $n$ is a principal quantum number. This solution is not quantized in a
strict sense. The energy of the atomic state depends on the particular value
of the parameter $a$ which can vary continuously. It is reasonable to look
for those $a$ values for which energy approaches its lowest value at a fixed
nuclear charge $Z$ and quantum number $n$. This analysis leads us to the
value for $a=1$ and we immediately arrive at the solution obtained by Niels
Bohr in the ''old'' quantum theory, namely $E_1(1,1)=-9/16=-$ $0.\,5625$
a.u. for the H negative ion and $E_1(1,2)=-49/16=$ $-3.\,0625$ a.u. for He.
These numerical results predict much lower ground state energies for these
two-electron atoms, than it is experimentally observed, see Table 1. For
this a simple qualitative explanation can be found. Above, we considered a
case when $a$ exactly equals $1$ -- or, in otherwords -- electrons are
constantly and exactly on the opposite ends of a diameter of a circular
orbit. Obviously for the low lying states of atoms this assumption
contradicts the uncertainty principle. If we take uncertainty relations into
account it means that we can not say that we can permanently know the exact
location of one electron with respect to another one. As a result one can
not assure that both electrons occupy all the time this most favourable from
the view-point of energy spatial configuration and, as a result, the actual
anergy level is raised in comparison with this most favourable classical
state.

\section{Ground state energy levels of a two-electron atom -- simple
expression}

For a single electron atom the general solution of the Schroedinger equation
is $E_n^{\left( s\right) }(Z)=Z^2/\left( 2n^2\right) $ we can express
ionization energy for the two-electron atom in our model as 
\begin{equation}
I\left( a,Z\right) =-\frac{\left( 1+a\right) ^2}{2(1+a^2)}\left( Z-\frac
a{(1+a)^2}\right) ^2-\frac{Z^2}2.  \label{eq3}
\end{equation}
Now in the framework of this model we can analyze what is the smallest
nuclear charge $Z_0$ for which stabile two electron negative ion will still
exist. Moreover, for this numerical exercise, let us now let $Z$ vary
continuously and let us analyze the most favourable -- with lowest energy --
configuration when $a=1.$ The threshold value of $Z$ at which ionization
energy $I\left( a,Z\right) $ approaches zero and consequently one electron
can be ejected form the atom, which means that the Helium-like ion ceases to
exist is 
\begin{equation}
Z_0=\frac 12+\sqrt{\frac 18}\approx 0.85355.  \label{eq4}
\end{equation}
It can be interpreted in a sense that at $Z=Z_0$ only those hypothetical
negative ions for which both electrons are \emph{all the time} \emph{exactly}
on the opposite ends of the diameter of a circular orbit $\left( a\equiv
1\right) $ can survive. This means that in this limit of small $Z$ values
the expression 
\begin{equation}
E_1\left( a=1,Z\propto 1\right) =-\left( Z-1/4\right) ^2  \label{eq4a}
\end{equation}
can be a good approximation for the energy levels for this very fragile
hypothetical Helium like ion. We realize very well that the last analysis is
only a numerical exercise performed with Eq. $\left( \ref{eq3}\right) $ and
in a real world there do not exist ions with fractional nuclear charge $Z$,
but, on the other hand, we will show further, that the obtained ground state
energy expression $\left( \ref{eq4a}\right) $ can be successfully used as
asymptotic approximation for ions with small integer $Z$ values.

Probably, such a hypothetical ion with $Z_0\approx 0.85355$ if it existed
could not survive for very long before autoionizaton, because, as it was
mentioned above, the assumption about permanent location of the electrons at
opposite ends of the diameter contradicts the uncertainty principle.

On the other hand it is known that the variational principle with only one
variational parameter -- the effective nuclear charge -- can lead to the
expression for energy in the form \cite{be1} 
\begin{equation}
E_1^{\left( v\right) }\left( Z\right) =-\left( Z-\frac 5{16}\right) ^2.
\label{eq5}
\end{equation}
This formula is known to give rather good agreement with experiment at the
opposite limit when $Z\gg 1$. This fact, namely that formula $\left( \ref
{eq5}\right) $ is good at the limit of large $Z$ values, can be easily
understood.

The classical text book problem is to calculate the energy states of a
two-electron atom by means of perturbation theory. At the very beginning one
can easily solve the problem neglecting electron -- electron interaction
(problem of independent electrons)\ and than one can consider electron --
electron interaction as a perturbation. Obviously, this approach will not be
very good for small $Z$ values, when nucleus -- electron interaction and
electron -- electron interaction is of the same magnitude. But if $Z$ is
getting larger and larger, approximation is getting better and better, since
nucleus -- electron interaction is getting strictly predominant. In the
first order the perturbation theory gives for energy \cite{be1} 
\begin{equation}
E_1^{\left( p\right) }\left( Z>>1\right) =-Z^2+\frac 58Z  \label{eq5a}
\end{equation}
which in a large $Z$ limit coincides exactly with $\left( \ref{eq5}\right) $.

Let us now combine the expression for small $Z$ values -- Eq. $\left( \ref
{eq4a}\right) $ and that for large $Z$ values -- Eq. $\left( \ref{eq5}%
\right) $ into one. Such an expression can be constructed in a rather simple
form

\begin{equation}
E_1\left( a,Z\right) =-\left( Z-\frac{1+\frac 14\sqrt{1-Z_0/Z}}4\right) ^2.
\label{eq6}
\end{equation}
The energies given by this expression can be compared with ground state
energies obtained for a two-electron atom with nuclear charge $Z$. The first
method for calculating these energies was worked out by Hylleraas \cite{hy1}%
. The method was based on a variation approach combined with perturbation
theory. If function with 50 summands was used as a probe function Hylleraas
obtained after laborious calculations ground state energy dependence on $Z$
for a two-electron atom in the form \cite{hy2} 
\begin{equation}
E_1^{(H)}\left( Z\right) =-Z^2+\frac 58Z-0.157652+O\left( \frac 1Z\right) .
\label{eq7}
\end{equation}

Finally, the quality of both $Z$ energy dependences must be compared with
experimental results for two electron atoms. The best experimental values
for these energies for wide range of ions can be found in the National
Institute of Standards data base \cite{bs1}.

\section{Discussion}

Today advanced numerical calculations of two electron atoms are available.
For the classical example of the ground state of a helium atom the
nonrelativistic energy of the ground state is obtained with an accuracy of
one part in $10^{19}$. For this a basis set containing $2114$ terms was
used, see \cite{da1} and references therein.

Despite these spectacular achievements it is still interesting, in our
opinion, to analyze a simple and easy understandable approach to helium-like
atoms. The experimental values of ground state energies for He-like atoms,
together with energies given by expressions Eq. (\ref{eq6}) and (\ref{eq7})
are collected in Table 1. In the third column of this table experimental
values of ground state energies of He-type atoms are given. The fourth
column contains the energies for the same ions calculated from Eq. (\ref{eq6}%
), as derived in this work. The fifth column of Table 1 contains energy
values calculated from the Hylleraas expression (\ref{eq7}) obtained by
means of a variational approach combined with the perturbation theory.
Energies from both expressions agree relatively well with each other for all 
$Z$ values, and with experimental data for small $Z$. For larger $Z$
formulae yield larger energies (less negative) than those measured in
experiment. This difference is due to relativistic and QED corrections that
are not included in these formulas. During the years relativistic and QED
corrections have been measured and calculated many times with increasing
accuracy, see for example \cite{qed1}. Nevertheless, for our purpose, when
we do not expect spectroscopic accuracy for the simple formulae under
discussion, an estimate of these effects can be used to account for them.
For an arbitrary two-electron atom the relativistic correction to the ground
state energy can be calculated as \cite{be1} 
\begin{equation}
E_{rel}=-\frac 18\alpha ^2Z^2(Z^2-3.606Z+3.29+0.05Z^{-1})  \label{eq8_1}
\end{equation}

In a similar way the QED\ correction for ground state energy can be found as%
\cite{be1} 
\begin{eqnarray}
E_{QED} &=&\frac{16Z^4\alpha ^3}{6\pi }\left[ \left( 3.745-\ln Z\right)
-Z^{-1}\left( 5.97-1.31\ln Z\right) +\right.  \nonumber \\
&&\left. Z^{-2}\left( 3.08-0.28\ln Z\right) \right]  \label{eq9_1}
\end{eqnarray}
These corrections must be added to the ground state energy values obtained
from expressions Eq. $\left( \ref{eq6}\right) $ and Eq. $\left( \ref{eq7}%
\right) $. If we now compare the corrected ground state energies obtained
with experimental values, agreement is very good. In Figure 1 the relative
differences $(E_e-E_t)/E_e$ between theoretical energy $E_t$ calculated from
the respective formulae $\left( \ref{eq6}\right) $ or $\left( \ref{eq7}%
\right) $, together with corrections $\left( \ref{eq8_1}\right) $ and $%
\left( \ref{eq9_1}\right) $, and experimental energy $E_e$ are presented. We
can see that in most cases the relative difference is less and in most cases
significantly less than $0.1\%$. An exception is the energy from the
Hylleraas formula for the negative H ion. These discrepancies can not be, at
least not significantly, reduced if we take into consideration the finite
mass of the nucleus. An account for this can lead to the relative increase
of the ground state energy in comparison of the value obtained from $\left( 
\ref{eq7}\right) $ for about $m/M$ where $m$ is a mass of the electron, but $%
M$ is the mass of the nucleus \cite{be1}. Apart from this simple effect of
the finite mass of the nucleus upon the energy levels of the ion, there
exist secondary -- usually smaller -- effects connected with correlation in
electron motion\cite{be1}. Nevertheless, it is known that correlation
effects in electron motion for a two-electron atom in quantum mechanics are
insignificantly small for ground state $S$ energy levels, and this
correction for ground state energy can be neglected.

Account for finite nucleus mass in Bohr's model when both electrons are
constantly located at the opposite ends of a diameter of a circular orbit
can differ from usual conclusions about the atomic $S$ state. In this model
the motion of both electrons is obviously strongly correlated. It must
strongly decrease the influence of nuclear motion upon the calculated ground
state energy, if not cancel it totally. In the case, when both electrons are
all the time on a circular orbit on opposite ends of a diameter, the nucleus
will remain at rest all the time in laboratory coordinates.

In the overall, approximation given by Eq. $\left( \ref{eq6}\right) $, which
was obtained practically without any adjustable parameters agrees with
experiment better than the Hylleraas formula. Maybe this can be considered
only as a curios coincidence, but we think that this expression for ground
state energy of a two-electron atom may be of some interest. As a starting
point for the derivation of the energy for a He-like atom ground state the
approach of old Bohr's quantum mechanics was used. This approach in some
textbooks still is viable for achieving intuitive understanding of a theory
of atomic structure.

One may ask why the square root $\sqrt{1-Z_0/Z}$ was chosen in formula $%
\left( \ref{eq6}\right) $. Any other power of $p$ in the expression $\left(
1-Z_0/Z\right) ^p$ would give the same asymptotic behaviour of the energy
for small as well as for large $Z$ values. The actual power $p=1/2$ was
chosen on the ground of the best coincidence with experimental results,
combined with the willingness to obtain a simple final expression. Actually,
the best coincidence with experimental results can be obtained at $p$ value
slightly larger than $1/2$ (around $0.56$). One must realize, that an exact
value of a parameter $p$, at which the best coincidence between experiment
and expression $\left( \ref{eq6}\right) $ can be obtained, varies slightly
for different $Z$ values. Nevertheless, it is surprising how close it
remains to $0.56$ in a broad range of $Z$ values. Taking into account all
this and the obvious circumstance, that one can not expect very high
accuracy from such a simple expression as $\left( \ref{eq6}\right) $ which
is not based on very sound assumptions, and trying to derive a simple
expression we have chosen $p=1/2$ in a final expression.

\section*{Table caption}

\textbf{Table 1.} Experimental values of the ground state energies for a
He-like atoms \cite{bs1}, the energies obtained from Eq. $\left( \ref{eq6}%
\right) $ and Eq. $\left( \ref{eq7}\right) $ together with the relativistic
and QED corrections to these energies. Energy is given in atomic units.

\section*{Figure caption}

\textbf{Figure 1.} The relative differences between the corrected ground
state energies for a He-type atoms calculated from Eq. $\left( \ref{eq6}%
\right) $ (squares) and Eq. $\left( \ref{eq7}\right) $ $(circles)$ and
experimental values for these energies.

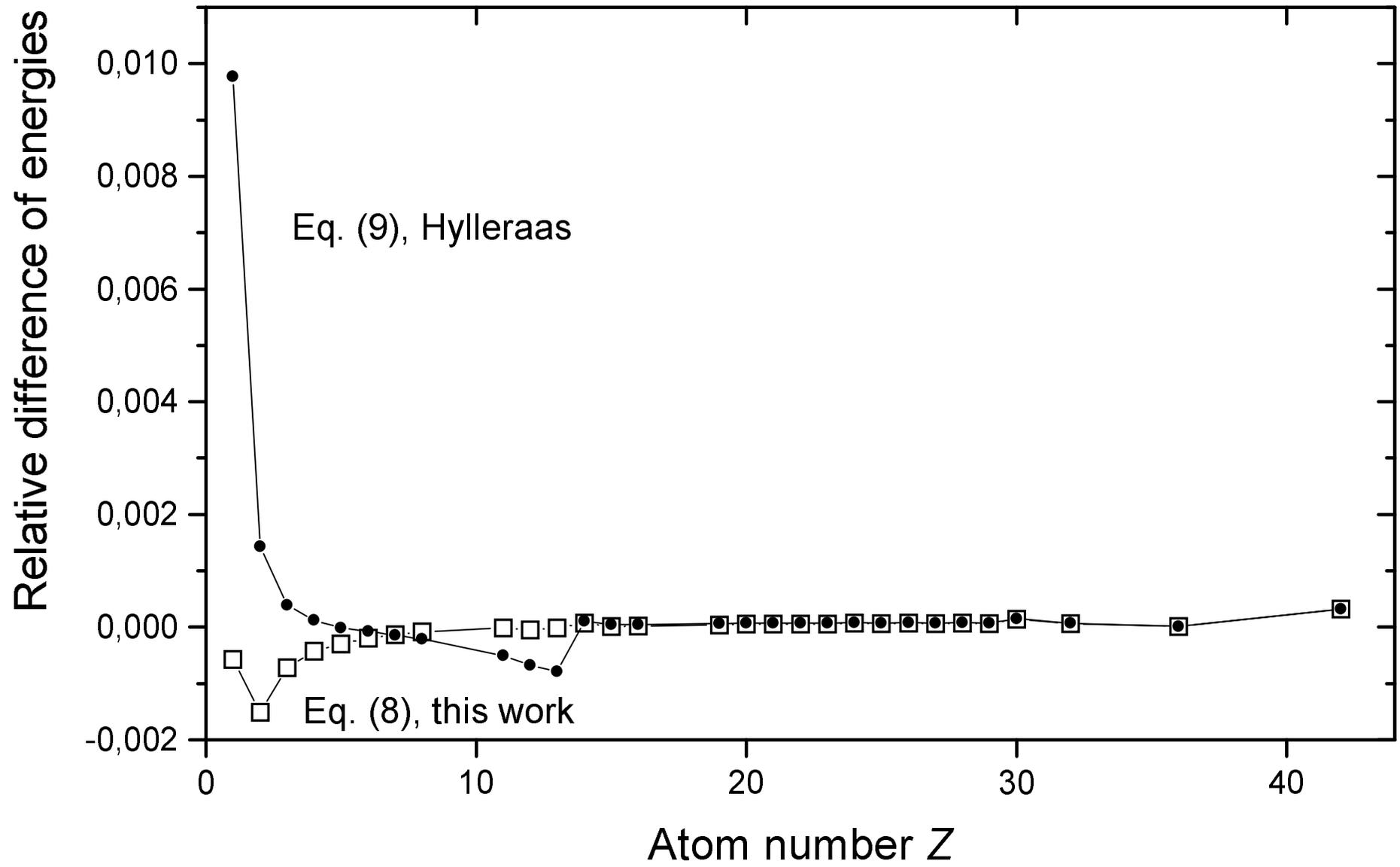
\end{document}